\documentstyle[preprint,aps,epsf]{revtex}

\begin{document}

\draft

\preprint{HIP-1998-08/TH}

\title{Limits on $\tan \beta$ in SU(5) GUTs with Gauge-Mediated
Supersymmetry Breaking}

\author{H.~Hamidian$^{\rm a}$, K.~Huitu$^{\rm b}$,
K.~Puolam\"{a}ki$^{\rm b}$, and D.-X.~Zhang$^{\rm b}$}

\address{$^{\text{a}}$Department of Physics, Stockholm University, Box
6730, S-113~85 Stockholm, Sweden}

\address{$^{\text{b}}$Helsinki Institute of Physics, P.O.Box 9,
FIN-00014 University of Helsinki, Finland}

\maketitle

\begin{abstract}

By considering the constraints from nucleon decay we obtain upper
limits on $\tan \beta$ in generalized supersymmetric
$SU(5)$ grand unified theories with gauge-mediated supersymmetry
breaking. We find that the predicted values of $\tan
\beta$ in these models are mostly inconsistent with the 
constraints from nucleon decay.
\\

\end{abstract} 


Supersymmetric (SUSY) grand unified theories (GUTs) are presently
considered to be among the most promising candidates for physics
beyond the standard model (SM).  However, for phenomenological
reasons, supersymmetry cannot be exact and it is usually assumed that
the theory includes a visible sector containing the observable
particles, and a hidden sector where supersymmetry is broken.  SUSY
breaking can be communicated to the visible sector either by
gravitational interactions, as in supergravity (SUGRA) inspired
models, or by SM gauge interactions, as in theories with
gauge-mediated SUSY breaking (GMSB).

GMSB models were initially studied in the early 1980's \cite{gmsb} and
have recently become the subject of much theoretical investigation. The
revival of interest in these models \cite{dineetal} is largely due to
recent dramatic improvements in our understanding of nonperturbative
effects in SUSY gauge theories as a result of the pioneering works of
Seiberg \cite{seiberg1} and Seiberg and Witten \cite{seiberg2}. Many
new mechanisms for dynamical SUSY breaking (DSB) have been found since
the appearance of \cite{seiberg1,seiberg2} and new DSB models have
been constructed \cite{skiba}. From a phenomenological point of view,
GMSB theories are interesting for a number of reasons. In these
theories gauge interactions provide flavor-symmetric SUSY breaking
terms and thus naturally suppress the flavor-changing neutral currents
associated with soft squark and slepton masses. They also predict
approximately degenerate squark and slepton masses (see below) and are
specified by a relatively small number of parameters.

In GMSB theories the $SU(3) \times SU(2) \times U(1)$ gauge
interactions of the ``messenger'' fields communicate SUSY breaking
from a hidden sector to the fields of the visible world.\footnote{Here
we do not consider the modification of direct mediation models such as the one proposed, e.g., 
in~\cite{comment1}.}  In the simplest of such
models \cite{dineetal}, in addition to the particles in the minimal
supersymmetric standard model (MSSM), there exists at least one
singlet superfield $S$ which couples to vector-like messenger
superfields $V+\overline V$ through the superpotential interaction
\begin{eqnarray}
W_{\rm mess}=\lambda_V SV\overline V.
\label{eq1}
\end{eqnarray}
At a scale $\Lambda\sim 10-100$ TeV, which is not much higher
than the weak scale, SUSY is broken and both the lowest and
$F$-component, $F_S$, of the singlet superfield $S$ acquire vacuum
expectation values (VEVs) through their interactions with the hidden
sector.  The VEV, $\langle S \rangle$, gives masses to the vector-like
supermultiplets $V+\overline V$, while $\langle F_S \rangle$ induces
mass splittings within the supermultiplets. As a result, the gaugino
and sfermion masses are generated through their gauge couplings to the
messenger fields. The gauginos receive masses at one-loop,
$m_{\lambda} \sim (\alpha / 4\pi) \Lambda$, where $\Lambda=\langle F_S
\rangle /\langle S\rangle$, while squarks and sleptons
do so only at two-loop order, ${\tilde m}^2 \sim (\alpha / 4\pi)^2
\Lambda^2$. This implies that $m_{\lambda} \sim {\tilde m}$, which is
one of the attractive features of GMSB theories.

In the minimal version of the GMSB model \cite{dineetal}, the
messenger fields belong to the ${\bf 5}+\overline {\bf 5}$ or ${\bf
10}+\overline {\bf 10}$ representations of the $SU(5)$ gauge group,
and the messenger Yukawa couplings, $\lambda_V$'s in (\ref{eq1}), in
any given $SU(5)$ representation are taken to be equal at the
unification scale $M_{\rm GUT}$.  Consequently, the spectrum at the
messenger scale consists of a set of fields in complete SU(5)
representations and the mass splitting among the fields in a
representation is induced through the renormalization group running of
the messenger Yukawa couplings from $M_{\rm GUT}$ down to the
messenger scale $\Lambda_{\rm m}$.

While the phenomenological implications of the minimal GMSB model have
been extensively studied \cite{gmsbph,borzu,carone}, non-minimal
generalizations of this class of theories have seen less
investigation. In this Letter, we study generalized models of GMSB
proposed by Martin \cite{martin} in which the messenger fields do not
necessarily form complete $SU(5)$ GUT multiplets.  It is, in fact, not
difficult to see how one may be naturally led to consider messenger
fields which belong to incomplete representations of the $SU(5)$ gauge
group in the generalized GMSB models. This is because the unification
of the messenger Yukawa couplings at the GUT scale---whose MSSM
analogue is the so-called $b-\tau$ unification \cite{chanowitz}---is
not necessarily required for gauge unification.  Suppose, for example,
that in addition to $S$ there exist singlet superfields, $S'$, whose
VEVs (but not the VEVs of their $F$-components) are just below the GUT
scale, and which couple only to some components of an $SU(5)$
multiplet\footnote{ Horizontal symmetries, e.g., can be used to
construct such models \cite{Nir:1993mx}.}. Then within the $SU(5)$
multiplet these superfields acquire masses of order ${\cal O}(M_{\rm
GUT})$ and decouple from the low-energy spectrum. The other
components, which get their masses only through couplings with the
superfield $S$, obtain masses of order $\lambda \langle S\rangle 
\sim \Lambda_{\rm m}$. Since $\sqrt{F_S}$ is much smaller than the 
masses of the heavy
superfields, these (missing) particles make negligible mass
contributions and play a less important role in determining the MSSM
mass spectrum.
    
A fruitful approach for examining the phenomenological viability of
SUSY GUTs has been to study processes that contribute to the nucleon
decay \cite{arnowitt}. However, in contrast to the minimal GMSB
theories, in which constraints from the nucleon decay yield quite
strong results \cite{carone}, the situation in this regard is somewhat
more complicated in the generalized GMSB models. In the minimal model
one begins with the renormalization group running of the SM gauge
coupling constants from the electroweak scale up to the GUT scale to
determine the mismatch between the SM and the $SU(5)$ gauge coupling
constants at the GUT scale. The mismatch, which is expressed in terms
of the sum of the contributions coming from threshold corrections at
the weak scale, the messenger scale, and the GUT scale, can then be
used to calculate the masses of the color-triplet Higgs bosons,
$M_{H_C}$.  On the other hand, in the generalized GMSB models the mass
splitting within a given $SU(5)$ representation is undetermined and
may result in large contributions to the mismatch at the GUT scale.
This means that the masses of the color-triplet Higgs bosons cannot,
in general, be reliably determined in the same way as in the minimal
model.

However, it is still possible to obtain useful constraints in the
generalized $SU(5)$ GMSB models.  The color-triplet Higgs superfields
belong to the ${\bf 5}+\overline{\bf 5}$ representation of $SU(5)$
and, in general, the unification scale in $SU(5)$ GUTs can be reliably
taken to be $M_{\rm GUT} \sim 2 \times 10^{16}$ GeV
\cite{carone}. This is due to the fact that above the GUT scale,
e.g. at around the reduced Planck scale, $2.4 \times 10^{18}$ GeV,
other effects---such as those coming from string theory, for
example---are expected to play an essential role. It is then safe to
conclude that the masses of the color-triplet Higgs fields cannot be
much larger
than the GUT scale.  As a conservative
upper bound, one can take $M_{H_C} \leq 10^{17}$ GeV.
[Note that in the minimal GMSB model the calculated masses of
the color-triplet Higgs bosons are found to be around
$10^{15}-10^{16}$ GeV if the LEP result $\alpha_3(m_Z) = 0.116 \pm
0.005$ is used \cite{erler}.]


Let us now describe the GMSB models that are the subject of this Letter.
Following Martin \cite{martin}, we shall consider five possible types of (chiral) superfields in the 
messenger sector of the generalized GMSB model
\begin{eqnarray}
n_L:~~~~ && L+{\overline L}=({\bf 1},{\bf 2},-\frac{1}{2}) + {\rm
conj.},\nonumber\\ n_D:~~~~ && D+{\overline D}=({\overline {\bf
3}},{\bf 1},\frac{1}{3}) + {\rm conj.},\nonumber\\ n_E:~~~~ &&
E+{\overline E}=({\bf 1},{\bf 1},1) + {\rm
conj.}\nonumber\\ n_U:~~~~ && U+{\overline U}=({\overline {\bf 3}},{\bf
1},-\frac{2}{3}) + {\rm conj.},\nonumber\\ n_Q:~~~~ && Q+{\overline
Q}=({\bf 3},{\bf 2},\frac{1}{6}) + {\rm conj.},
\label{eq:mfields}
\end{eqnarray}
where the multiplicities of the messenger fields are denoted by $(n_L,n_D,n_E,n_U,n_Q)$.

Requiring that the gauge couplings remain perturbative, and assuming
messenger field masses that do not greatly exceed $10^4$ TeV, leads
(see \cite{martin} for further discussion) to the following set of
multiplicites for the messenger fields
\begin{eqnarray}
(n_L,n_D,n_E,n_U,n_Q) \leq & (1,2,2,0,1)\nonumber\\ {\rm or}~ &
(1,1,1,1,1)\nonumber\\ {\rm or}~ & 
(1,0,0,2,1)\nonumber\\ {\rm or}~ &
(4,4,0,0,0).
\label{eq:martinlimits}
\end{eqnarray}

The general low energy superpotential of the messenger sector is
\begin{eqnarray}
W_{\rm mess} =\sum_{n_L} \lambda_L^i SL^i\overline L^i + \sum_{n_D}
\lambda_D^i SD^i\overline D^i ~~~~~~~~~~~~~~~~~~~~&\nonumber\\
+\sum_{n_E} \lambda_E^i SE^i{\overline E}^i +\sum_{n_U} \lambda_U^i
SU^i{\overline U}^i +\sum_{n_Q} \lambda_Q^i SQ^i{\overline Q}^i&.,
\end{eqnarray}
and the MSSM spectrum can be determined once the messenger sector is
fixed.  In the numerical estimates that are reported here, we shall
use $\Lambda_{\rm m}=10^4$ TeV.

We can now directly calculate the nucleon decay rates for the GMSB
models listed above by using the process $n \rightarrow K^0 {\overline
\nu}_\mu$ as the characteristic mode. The short- and long-distance
corrections and the hadronic matrix elements are taken at their
conservative values, as e.g. in \cite{carone}, giving a lower bound on
the product $M_{H_C} \sin 2 \beta$. We have studied all the 53
possible models which satisfy criteria (\ref{eq:martinlimits}) and
contain massive gauginos. By setting the upper limit on the mass of
the triplet Higgs at $10^{17}$ GeV, upper bounds on $\tan \beta$ can
be obtained and some interesting configurations for $\Lambda=\langle
F_S \rangle /\langle S\rangle=100$ TeV are listed in
Table~\ref{tab:cases}. 
The parameters in the columns are the ones needed in the nucleon
decay formula.
We find a general bound, $\tan \beta < 10$,
except in cases (1,4,0,0,0), (2,4,0,0,0), and (1,3,0,0,0) for which
this bound is $\tan \beta < 17$.

There is also a lower bound on $\tan \beta$ following from the nucleon
decay constraints. However, this constraint is less severe than
the one
obtained by requiring that the Yukawa couplings should not blow up at
the GUT scale, giving $\tan \beta > 0.85$ \cite{carone}.

For comparison, we have also calculated the value of $\tan \beta$ with
the assumption of radiatively-broken $SU(2) \times U(1)$ symmetry when
trilinear and bilinear soft couplings vanish at the messenger scale---which, among other things, free 
these models from the
supersymmetric CP problem and make them extremely predictive.
For the computations we use the full one-loop effective potential~\cite{borzu}.
The calculated values of $\tan \beta$ are plotted versus the upper limit from
nucleon decay in Fig.~1. 
The calculations are done for $\Lambda =100$ TeV (black circles) and 
$\Lambda =200$ TeV (open rectangles).
For $\Lambda =100$ TeV, the values of $\tan\beta $ are all larger than
bounds from nucleon decay, whereas
for $\Lambda=200$ TeV, the values of $\tan\beta$ from radiative symmetry 
breaking
roughly double, and two of the models are allowed.
These are the first two in Table 1.
Note that less than 53 points are plotted in Fig.~1 since 
many of these models are on top of each
and some of them (four models) turn out not to be 
physical.

To conclude, we have studied the phenomenological viability of generalized 
supersymmetric
$SU(5)$ grand unified theories with gauge-mediated SUSY breaking 
by calculating the upper limits on $\tan \beta$ from nucleon decay
in these theories. We find that the predicted values of $\tan \beta$  
are mostly  
inconsistent with the constraints from nucleon decay. 
Our results suggest that if theories with GMSB are to be taken as serious 
SUSY GUT candidates 
beyond the standard model, 
the bilinear and/or trilinear soft terms cannot vanish at the messenger
scale, and gauge groups other than $SU(5)$---together with their associated 
implementation of dynamical SUSY breaking---are required for acceptable 
low-energy phenomenology.

\section*{Acknowledgements}

HH thanks the Swedish Natural Science Research Council for financial
support.  The work of KP is supported by a grant from the Magnus
Ehrnrooth Foundation, and the work of KH, KP and DXZ by the Academy of
Finland (No.~37599).

\newpage

\begin{table}
\caption{Limits on $\tan \beta$ with $\Lambda =100$ TeV.}
\label{tab:cases}
\begin{tabular}{crrrrrr}
$(n_L,n_D,n_E,n_U,n_Q)$\tablenotemark[1] & $\left( \tan \beta
\right)_{\text{max}}$\tablenotemark[2] &
$\frac{m_{\text{Bino}}}{\text{GeV}}$ &
$\frac{m_{\text{Wino}}}{\text{GeV}}$ &
$\frac{m_{\text{Gluino}}}{\text{GeV}}$ &
$\frac{m_{\tilde{q}}}{\text{GeV}}$ &
$\frac{m_{\tilde{e}}}{\text{GeV}}$ \\ \tableline
$(1,4,0,0,0)$\tablenotemark[3] & 17 & 301 & 263 & 2565 & 2262 & 373 \\
$(1,3,0,0,0)$\tablenotemark[3] &13 & 245 & 263 & 1971 & 1894 & 369 \\
$(2,4,0,0,0)$\tablenotemark[3] & 11 & 384 & 521 & 2565 & 2287 & 527 \\
$(1,1,0,0,0)$\tablenotemark[4] & 5 & 301 & 263 & 724 & 1006 & 383 \\
\end{tabular}
\tablenotetext[1]{$n_k$ specify the numbers of different types of 
messenger fields as defined in eq.~(\ref{eq:mfields}).}
\tablenotetext[2]{Limit assumes $M_{H_C} \le 10^{17} \text{GeV}$.}
\tablenotetext[3]{These three configurations are the only choices
which allow $\tan \beta >10$.}  \tablenotetext[4]{$5+{\overline{5}}$
model.}
\end{table}

\begin{figure}
\caption{The upper limit on $\tan \beta$ from nucleon decay vs. 
$\tan\beta$ calculated by assuming radiatively-broken $SU(2) \times U(1)$ 
symmetry 
and vanishing bilinear and trilinear soft couplings at the messenger scale.
The black circles correspond to $\Lambda=100$ TeV and the open rectangles
to $\Lambda=200$ TeV.} 
\label{fig:tanplot}
\mbox{\epsfxsize=17cm \epsfysize=17cm \epsffile{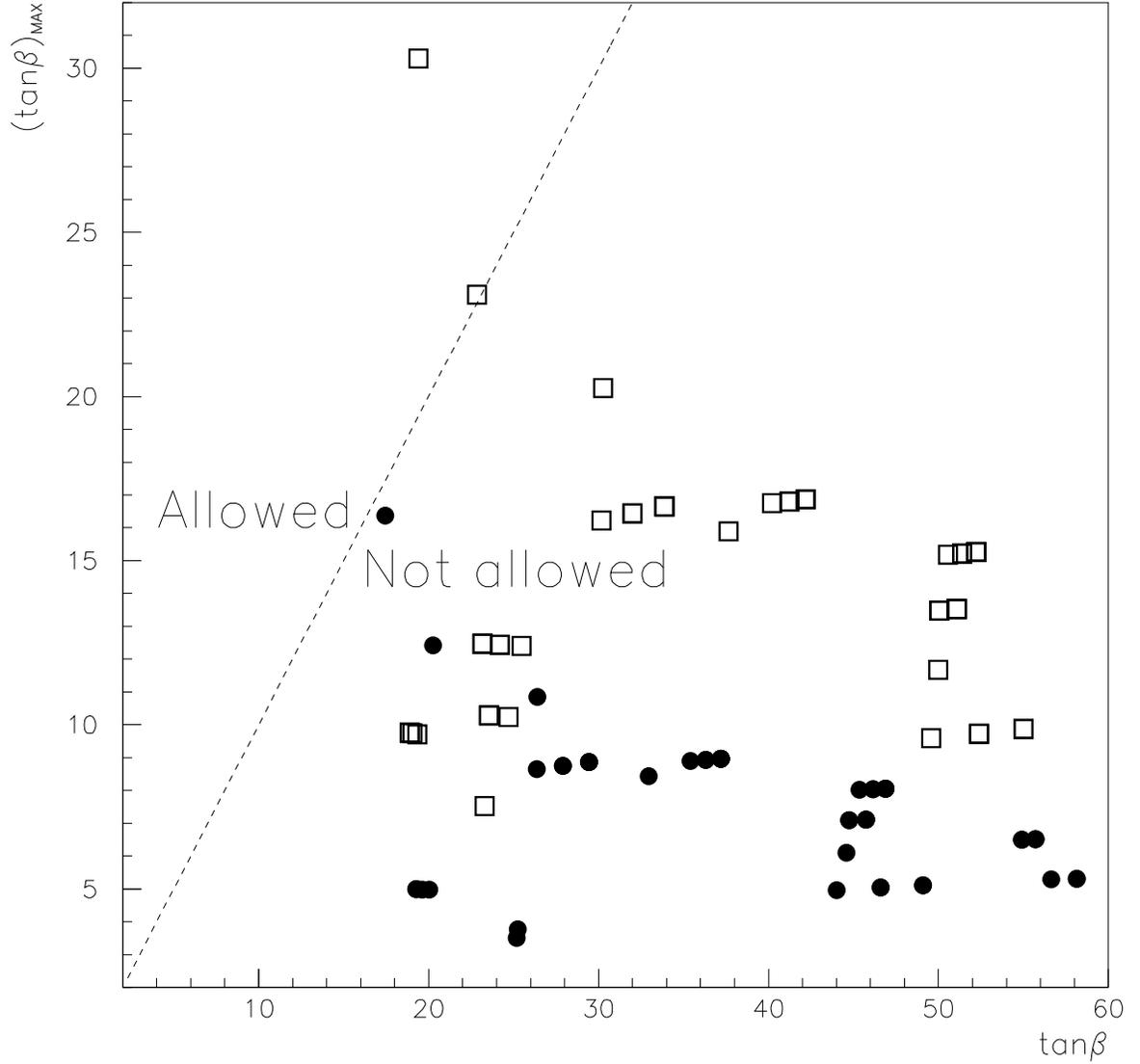}}
\end{figure}

\end{document}